\documentclass{article}
\pdfoutput=1
 \usepackage[final, nonatbib]{neurips_2021}
\usepackage{algorithm2e}
\usepackage{amsthm}
\usepackage{amsmath}
\usepackage{physics}
\DeclareMathOperator*{\argmax}{arg\,max}

\usepackage{float}
\usepackage{graphicx}

\usepackage[utf8]{inputenc} 
\usepackage[T1]{fontenc}    
\usepackage{hyperref}       
\usepackage{url}            
\usepackage{booktabs}       
\usepackage{amsfonts}       
\usepackage{nicefrac}       
\usepackage{microtype}      
\usepackage{mathtools}
\usepackage[table,xcdraw]{xcolor}
\usepackage{hyperref}
\hypersetup{
    colorlinks=true,
    linkcolor=blue,
    citecolor=blue,      
    urlcolor=blue,
    pdftitle={Overleaf Example},
    pdfpagemode=FullScreen,
    }
\usepackage{subcaption}
\usepackage{adjustbox}

\newcommand\blfootnote[1]{%
  \begingroup
  \renewcommand\thefootnote{}\footnote{#1}%
  \addtocounter{footnote}{-1}%
  \endgroup
}

\theoremstyle{definition}
\newtheorem{definition}{Definition}[section]
\title{Clarinet: A Music Retrieval System}

\author{
Kshitij Alwadhi$^\ast$\\
\textit{Dept. of Electrical Engineering} \\
\textit{Indian Institute of Technology Delhi}\\
   \And
    Rohan Sharma$^\ast$\\
\textit{Dept. of Electrical Engineering} \\
\textit{Indian Institute of Technology Delhi}\\
\And
Siddhant Sharma$^\ast$\\
\textit{Dept. of Electrical Engineering} \\
\textit{Indian Institute of Technology Delhi}
}

\begin{document}
\RestyleAlgo{ruled}
\SetKwComment{Comment}{/* }{ */}
\maketitle
\blfootnote{$^\ast$ Equal Contribution.}

\begin{abstract}


A MIDI based approach for music recognition is proposed and implemented in this paper. Our Clarinet music retrieval system is designed to search piano MIDI files with high recall and speed. We design a novel melody extraction algorithm that improves recall results by more than 10\%. We also implement 3 algorithms for retrieval-two self designed (RSA Note and RSA Time), and a modified version of the Mongeau Sankoff Algorithm. Algorithms to achieve tempo and scale invariance are also discussed in this paper. The paper also contains detailed experimentation and benchmarks with four different metrics. Clarinet achieves recall scores of more than 94\%.
All of our code is publicly available \href{https://github.com/rohans0509/Clarinet}{here}.

\end{abstract}

\section{Introduction}
Traditional Music recognition systems like Shazam and DejaVu rely on spectrogram analysis of samples to match songs. A major flaw in these algorithms is that they assume that the query given to the system has features (like sampling frequency) \emph{precisely} similar to the studio recorded versions. Thus, matching song derivatives (like instrumental covers) is not possible. 
\\
To solve this problem, we need to find derivative invariant features. One such feature is the melody of the song. Derivatives of the same song are very likely to have the melody intact. Thus, if we could detect this melody, it would make music recognition more general by encompassing song derivatives.
\\
This project aims to tackle this problem by developing a MIDI based retrieval system that relies on monophonic melody features. The system is tested and evaluated on the Maestro Dataset \cite{hawthorne2018enabling} which is a dataset of Piano songs.

\paragraph{Why MIDI?}
MIDI files record the audio's notes at any given time, giving a standardized interpretation of the song. A melody extracted as a midi can thus we read as sheet music. Much of music theory (like scale and tempo invariance) depends on standardised time signatures and note values. Hence, MIDI helps us understand and deal with purely musical content (as opposed to audio signal content)
\\The following section describes our problem statement. \textbf{Section 3} provides an in depth discussion of the Clarinet model. \textbf{Section 4} discusses Experiments and evaluation of results. In \textbf{Section 5} we summarise our results. \textbf{Section 6} details future work that can be done in this field and \textbf{Section 7} talks about possible applications of our work.
\newpage
\subsection{Contributions} 
In this paper, we state many results of independent importance and also tie the concepts together to create a end-to-end robust and fast music retrieval system. The following is a list of the results:
\begin{enumerate}
    \item \textbf{Melody Extractor}: Modified a state of the art melody extractor \cite{og_skyline2,og_skyline}, and improved it substantially both for independent use and for music retrieval. 
    \item \textbf{Similarity Computation}: Created RSA Note and RSA Time algorithms for music retrieval and modified a pre-existing technique Mongeau-Sankoff \cite{mongeau2,mongeau1} to make it more general.
    \item \textbf{Sliding Window}: Ideated and implemented a sliding window concept to match query audios with audios of larger duration, that allowed for more generalisation of pre-existing algorithms.
    \item \textbf{Evaluation Techniques}: Defined a metric called \textbf{Margin of Discrimination} that is the difference in confidence (similarity) scores between the target document and the next ranked document. This metric can be independently used in many applications.
    
\end{enumerate}
\section{Problem Statement}
This paper contains many independent algorithms that can be used for various use cases. Hence, we address various problem statements and requirements that are listed below:
\begin{enumerate}
    \item A need for a more \textbf{custom and robust melody extractor}, since the current state of the art method is too restrictive and generates unstable\footnote{Stability as defined in \textbf{Definition 3.1}} melodies.
    \item A \textbf{fast and accurate music retrieval system} that works when queries are of varying length.
    \item A methodology to ensure data is \textbf{tempo and scale invariant} to ensure generality of data.
\end{enumerate}
While we successfully tackle all the above prompts, we also touch upon other related ideas that have either been detailed in the paper itself or left as future work in \textbf{Section 6}.
\section{Clarinet Model}
\emph{Note that the Clarinet Model takes an input of a MIDI file. However, we also have added methodologies to convert MP3/WAV audio files to a Clarinet compatible format and back   \cite{omnizart}.}
\paragraph{Overview}
Clarinet, after inputting the MIDI file, clips the audio if required and then processes it\footnote{Audio is processed for the usage of Mongeau-Sankoff Algorithm and its derivatives only}. After the processing, the \textbf{melody is extracted}. For this, Clarinet provides two methods namely Skyline  \cite{purl,ozcan_melody_2005,og_skyline2,og_skyline} and the novel Modified Skyline. The following section contains the details on both the methods and compares them. Once we obtain the melody of the query, we can finally use it for \textbf{retrieving} the actual audio (which was pre-processed to contain the melodies as well). Clarinet provides three methods for the same, one of which is a modified version of the state of the art methods and the other two are novel algorithm dubbed RSA Time Algorithm and RSA Note Algorithm. All three methods use distinctly different ideas and have been detailed in subsequent sections. The RSA algorithms (RSA Time and RSA Note) are built on the principle of sliding windows of time and notes respectively and the Mongeau-Sankoff algorithm  \cite{mongeau2,mongeau1} is based on edit distance changes which is an extension of Levenshtein Distance.
\newpage
\begin{figure}[H]
	\centering
	\includegraphics[scale=1.2]{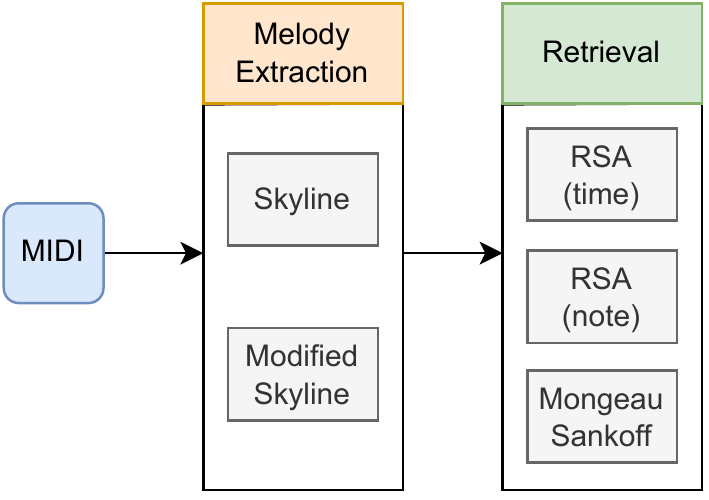}
				\caption{Pipeline of Clarinet}
\end{figure}

\subsection{Melody Extraction}
Raw MIDI files contain a lot of information of the audio. Extraction of the melody from the MIDI file is encouraged due to various reasons that are listed below:
\begin{enumerate}
    \item \textbf{Monophonic Audio} - All retrieval models detailed below in \textbf{Section 3.4.1-3.4.3} require the audio to be monophonic at all times due to the representation as given in \textbf{Section 3.2}.
    \item \textbf{Information Extraction} - The melodic element of the song is likely to be the same irrespective of the song derivative. Thus a majority of the relevant information of a song lies in its melody  \cite{herbert,pesek}.
    \item \textbf{Noise Reduction} - Extracting just the melody allows us to ignore the noise present in the audio file.
    \item \textbf{Data Compression} - Keeping only the melody of a song allows us to space taken to store our data.
    \item \textbf{Faster Retrieval} - Pruning the melody to be more precise also gives us the benefit of a faster solution.
    
\end{enumerate}
Let us now outline the current state of the art melody extraction algorithm, and the flaws with it in \textbf{Section 3.1.1}. After that in \textbf{Section 3.1.2}, we will detail the modified algorithm that takes care of the two prominent issues with the original skyline algorithm.

\subsubsection{Skyline Algorithm}
The \textbf{skyline algorithm}  \cite{purl,ozcan_melody_2005,og_skyline2,og_skyline} primarily works on the principal that the note with the \textbf{highest pitch constitutes the melody}. This is based on the assumption that the human ears tend to pick up on the higher frequencies and hence skyline uses these notes as the most prominent ones for the melody. The proposed algorithm is detailed below.
\newpage
\begin{algorithm}[H]
\caption{Skyline}\label{alg:Skyline}

\KwData{Raw Notes}
\KwResult{Melody Extracted Notes}
$notes \gets rawNotes$\;
$skylineNotes \gets []$\;
\For{$note \in notes$}{
    $sameTimeNote = \text{notes with same start as note}$\;
    \If{$note \notin \text{importantNotes } \&\& \text{ note =} \argmax \{pitch(sameTimeNote) \}$}{
        $nextNote \gets \text{next Note with different start time}$\;
        $note.end = \min (note.end,nextNote.start)$\;
        $\text{note appended to skylineNotes}$\;
    }
}
\end{algorithm}
And here we see that the skyline algorithm \textbf{loses on important information} by directly truncating and discarding notes without saving them for later. It also is \textbf{very restrictive} on the methodology of deciding the notes relevant to the melody of a song.
We develop a modified version of the skyline algorithm to address these short comings in the following section.
\vspace{-0.3em}
\subsubsection{Modified Skyline Algorithm}
\vspace{-0.2em}
The \textbf{modified version of skyline} perfects the original algorithm, while also adding \textbf{additional functionality} that replaces the current state of the art technology with ours. 
Working under the theory that only one note plays at any given time in the melody of the song, the Modified Skyline algorithm allows for a custom function that takes in the important note, and returns a real number that can be used to \textbf{compare between the importance of various notes}. 
Our preliminary implementation contains various criteria functions dependent on the pitch and velocity but this can be scaled up to any information the MIDI file contains.
Our implementation also deals with the unfortunate event where \textbf{notes are prematurely truncated in the original skyline algorithm} by saving the notes till their true end time.
\\
\begin{algorithm}[H]
\caption{Modified Skyline}\label{alg:Modified}
\KwData{Raw Notes}
\KwResult{Melody Extracted Notes}
$criteria \gets f: (note\to r\in\mathbb{R})$ \Comment*[r]{$r\in\mathbb{R}$ is the importance score of note} 
$notes \gets rawNotes$\;
$importantNotes \gets []$\;
$skylineNotes \gets []$\;
\For{$note \in notes$}{
    $sameTimeNote = \text{notes with same start as note}$\;
    \If{$note \notin \text{importantNotes } \&\& \text{ note =} \argmax \{criteria(sameTimeNote) \}$}{
        $\text{note appended to importantNotes}$\;
    }
}
\For{$note \in importantNotes$}{  
    $currentNotes \gets \text{All notes playing at note.start time}$\;
    $bestNote \gets \argmax criteria(currentNotes) $\;
    $bestNote.start \gets note.start$\;
    $bestNote.end \gets futureBestNote.start$\;
    $\text{bestNote appended to skylineNotes}$\;
}
\end{algorithm}
As can be seen from the above formulation of the two algorithms, our modification deals with two pertinent issues with the original skyline algorithm by implementing the following:
\vspace{-0.35em}
\begin{enumerate}
    \item Flexibility of defining one's own criteria for gauging importance of a note for comparison.
    \item Not losing information by prematurely truncating and deleting notes from hash.
\end{enumerate}
\vspace{-0.35em}
Another unexpected improvement was observed; the melody extracted was much more stable for the modified version of skyline when compared to the original version. 
\newpage
\subsubsection{MIDI Representation}
{\begin{definition}[\textbf{Stability of a melody}]
\emph{It is defined as the variation of the time a note plays compared to the average time that all notes play. Additionally, the lack of perturbations in the extracted melody's MIDI representation is also a measure of \textbf{stability of a melody}.}
\end{definition}}
Let us first look at the MIDI representation  \cite{MIDI_Rep} of the original audio.
\begin{figure}[H]
	\centering
	\includegraphics[width=.5\textwidth]{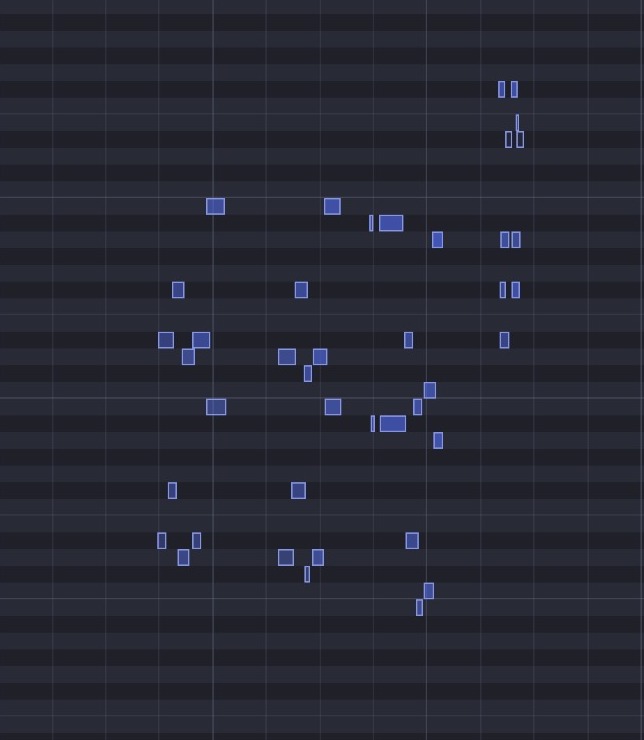}
				\caption{MIDI Representation of original audio}
\end{figure}
This representation contains the note being played plotted against the time stamp. We can see that at the same time, we sometimes have \textbf{multiple notes playing}. As discussed above, this is unnecessary and we need to extract the melody.
\\Now let us use the two melody extraction techniques and compare the results between the two.

\begin{figure*}[h]
    \centering
    \begin{subfigure}[t]{0.44\textwidth}
        \centering
        \includegraphics[height=2.2in]{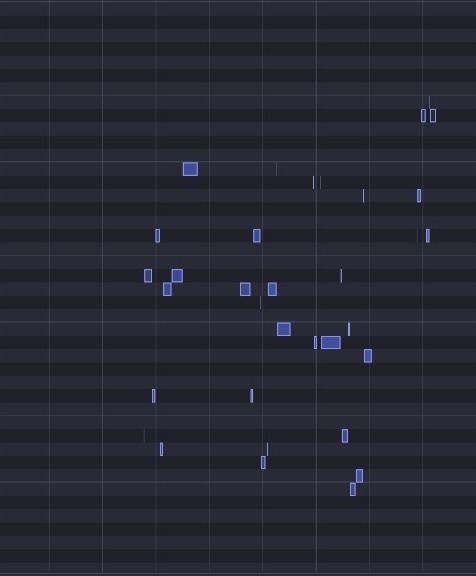}
        \caption{Skyline extracted melody}
    \end{subfigure}%
    ~ 
    \begin{subfigure}[t]{0.54\textwidth}
        \centering
        \includegraphics[height=2.2in]{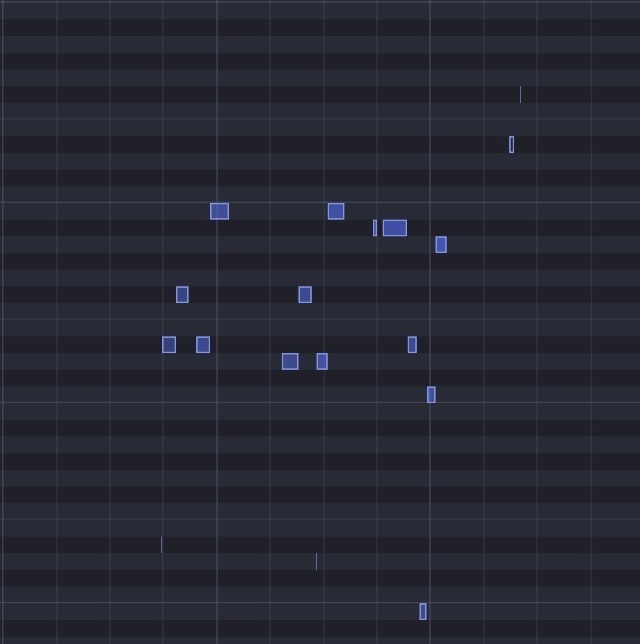}
        \caption{Modified skyline extracted melody}
    \end{subfigure}
    \caption{MIDI Representation on extracted melodies}
\end{figure*}
We see that the melody generated by original skyline algorithm (left) is very messy and contains anomalous notes being played as well. Clearly, this is quite poor in quality.
\\We can also see how much cleaner (and stable, as defined by  \textbf{Definition 3.1}) this modified skyline algorithm (right) really is, by observing the left representation against the right one.
\newpage
\subsection{Text Representation and Fuzzy Matching}
The approach we have taken involves representing the\textbf{ melody midi file as a string} and using \textbf{sequential matching algorithms} to retrieve similar results. The simplest way to convert midi files to text is to ascertain the \textbf{notes} played and apply a \textbf{boolean} matching algorithm.

\subsubsection{Naive Boolean Model}
The algorithm is as follows
\begin{enumerate}
\item \textbf{Text Representation}-Convert document and query MiDi file to string
\item \textbf{Boolean Matching}-Check if the query string is a \textbf{substring} of the document
\end{enumerate}
\textbf{\emph{Example}}-Suppose the Document  and query midi files are converted to text and obtained as 
\[D=ABCBGC\]
\[Q=CBG\]
Since \textbf{$Q\in D$}, the system returns a \textbf{match}.\\
Since the algorithm relies on substring matching, it is extremely \textbf{fast}. However, this method is too \textbf{prone to errors}. A slight change in the query notes from document notes will cause the system to fail.\\
\textbf{\emph{Example}}-Suppose the Document  and query midi files are converted to text and obtained as 
\textbf{\[D=ABCBGC\]}
\textbf{\[Q=ACBG\]}
Here the query contains another note \textbf{A}. This could be because the user played the note incorrectly or the midi conversion tool detected the wrong pitch. The system fails to find a match here since \textbf{$Q \notin D$} \\
The algorithm provides no ranking of similarity between the document and query. Thus,  we don't get ranked similarity results even when other songs in the database might be similar. We need  an algorithm more resilient to errors and one that provides similarity scores.
\subsection{Fuzzy Logic Model}
As discussed above, we need a string matching algorithm that doesn't give 0 scores for queries that don't \emph{exactly} match the document. Levenshtein distance (a kind of Dynamic Time Warping) has been shown empirically to be the best distance measure for
string editing. It measures how far two strings are using an operation set(insertion, deletion and substitution)

\begin{definition}[\textbf{Levenshtein Distance}]
\emph{The Levenshtein (string editing) Distance between two sequences is the minimal number of substitutions, insertions, and deletions needed
to transform from one to the other. Formally, the Levenshtein distance between
the prefixes of length i and j of sequences S and T, respectively, is:}

\[ 
lev_{s,t}(i,j)= \begin{cases*}
                    max(i,j) & if  min(i,j)=0  \\\\
                    min \begin{cases*}
                    lev_{s,t}(i-1,j)+1  \\
                    lev_{s,t}(i,j-1)+1  \\
                    lev_{s,t}(i-1,j)+(S_i\neq T_j)  \\
                 \end{cases*}
                 & otherwise
                 \end{cases*} 
\]%

\emph{Here we have assumed an \textbf{unweighted} Levenshtein Distance (Weights may also be introduced according the value of either of the edit operations). }
\end{definition}
\newpage
The above Levenshtein distance suffers from a major drawback. If the query and document are of different lengths, the edit distance becomes very large. Due to this large edit distance, \textbf{query variations will have no effect}-leading to poor discrimination. Since the queries we have chosen are only 10\% of the document length, the results will thus be very inaccurate.

\subsection{Similarity Calculation}
Based on the above discussion we arrive at the following requirements for a "good" model:
\begin{enumerate}
\item \textbf{Robustness}-The model should be able to handle noisy inputs. This includes queries having spurious additional notes ("ABDF" instead of "BDF"), substituted notes ("BGF" instead of "BDF") and fewer notes ("BF" instead of "BDF")
\item \textbf{Discrimination}-The model should be able to discriminate between documents effectively. In other words, the similarity scores of documents shouldn't be too close to each other. We have designed the Margin of Discrimination metric in Section 5 as a measure of this property.
\end{enumerate}

\subsubsection{RSA (Time Based)}
RSA is a \textbf{time based sequential retrieval} algorithm inspired by dynamic time warping. It assumes the approximate query time length is known beforehand and performs chunk based matching for subsets of the document. A detailed description of the algorithm is given below:
\\
Suppose the \textbf{approximate} query length is $T_0$.  For each query :
\begin{enumerate}
\item \textbf{Window creation}-Create a window of length $T_0$ over D, ie. pick the first $T_0$ seconds of D. Call this window $W$. Thus, we have $T_W=T_0$
\item \textbf{Distance Computation}-Since $T_Q\approx T_0 $ (from our assumption), we have $T_Q\approx T_W$. We can thus safely find the edit distance between Q and D. Store this as \textbf{$L_{Q,D}$} (standing for Levenshtein distance).
\item \textbf{Translation}-Translate the window by stride length (\textbf{$T_S$})  and compute the distance between $W_{new}$ and $Q$ (since we still have $T_Q\approx T_{Wnew}$)
\item \textbf{Compare distance}-Update \textbf{$L_{Q,D}$} if $W_{new}$ has smaller distance.
\item \textbf{Repeat}-Repeat step 3,4 till the document has no more windows
\end{enumerate}

From the above edit distance we can calculate the similarity as 
\begin{equation*}
    \boxed{sim(Q,D)=1-\frac{L_{Q,D}}{len(Q)}}
\end{equation*}

The similarity between the document and the query is thus the same as the best similarity between the set ($S=S_{T_0}$) of document substrings of length $T_0$.
\[sim(Q,D)=max(sim(Q,W))\]
\[W\in S\]
\newpage
\textbf{Example}

\begin{figure}[H]
	\centering
	\includegraphics[scale=1.0]{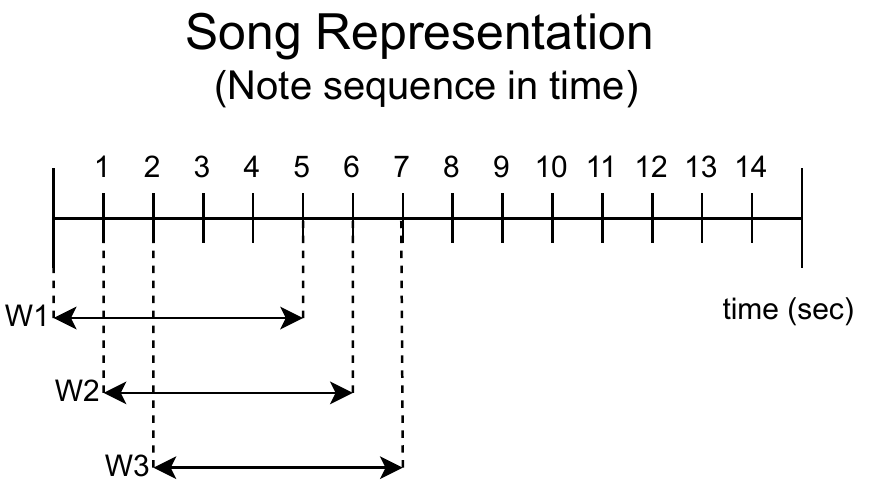}
				\caption{Working of RSA(Time)}
\end{figure}
In the above diagram we have \textbf{stride length}
$T_S=1s$
and \textbf{Approximate Query Length}
$T_0=5s$\\
Using this we can calculate the window size as
\[T_W=T_0=5s\]
Thus our window will be of 5s and will move 1s per iteration. The value of stride length is a hyperparameter. \emph{Lower stride lengths give higher accuracy but at the cost of time}.

\subsubsection{RSA (Note Based)}
Instead of looking at windows in the time domain, we can look at note windows. 
Suppose the \textbf{approximate} query length (in notes) is $N_Q$.  For each query :
\begin{enumerate}
\item \textbf{Window creation}-Create a window of length $N_Q$ over D, ie. pick the first $N_Q$ notes of D. Call this window $W$. Thus, we have $N_W=N_Q$. \emph{Note that here the window size depends on the query rather than being query invariant like the previous algorithm.}
\item \textbf{Distance Computation}-Find the edit distance between Q and W. Store this as \textbf{$L_{Q,D}$} (standing for Levenshtein distance).
\item \textbf{Translation}-Translate the window by stride length (\textbf{$N_S$})  and compute the distance between $W_{new}$ and $Q$ 
\item \textbf{Compare distance}-Update \textbf{$L_{Q,D}$} if $W_{new}$ has smaller distance.
\item \textbf{Repeat}-Repeat step 3,4 till the document has no more windows
\end{enumerate}

From the above edit distance we can calculate the similarity as 
\begin{equation*}
    \boxed{sim(Q,D)=1-\frac{L_{Q,D}}{len(Q)}}
\end{equation*}

The similarity between the document and the query is thus the same as the best similarity between the set ($S=S_{T_0}$) of document substrings of length $N_Q$ (if $N_S=1$).
\[sim(Q,D)=max(sim(Q,W))\]
\[W\in S\]
\newpage
\textbf{Example}
\begin{figure}[H]
	\centering
	\includegraphics[scale=1.0]{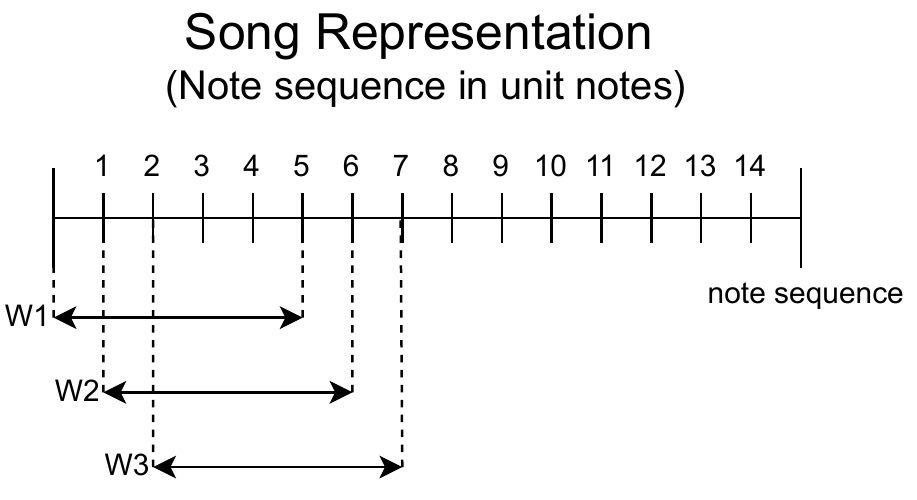}
				\caption{Working of RSA(Note)}
\end{figure}

In the above diagram we have \textbf{stride length}
$N_S=1$
and \textbf{Query Length}
$N_Q=5$\\
Using this we can calculate the window size as
\[N_W=N_Q=5\]
Thus our window will be of 5 \textbf{notes} and will move 1 \textbf{note} per iteration. The value of stride length is a hyperparameter. \emph{Again, Lower stride lengths give higher accuracy but at the cost of time}. For a given Query Q we can calculate the note window size as 
\[N_W=len(Q)\]
For reasonable accuracy, the stride length can then be chosen from
\[N_S\in [0,N_W]\]

\subsubsection{Mongeau Sankoff Retrieval}
The \textbf{Mongeau-Sankoff} algorithm is a retrieval algorithm designed specifically for \textbf{Music Retrieval}. It improves on the traditional unweighted Levenshtein Distance by adding pitch based substitution weights and accounts for duration of notes(which the above algorithms did not do). The Mongeau-Sankoff Distance between two midi files is given by

\[ 
ms_{s,t}(i,j)= 
                    min \begin{cases*}
ms_{s,t}(i-1,j-1) & $if   \alpha_i=\beta_j (match)$ \\ 
ms_{s,t}(i-1,j-1)+\delta(\alpha_i\rightarrow \beta_j) &   (substitution)\\
ms_{s,t}(i-1,j)+\delta(\alpha_i\rightarrow \epsilon) &    (deletion)\\
ms_{s,t}(i,j-1)+\delta(\epsilon \rightarrow \beta_j) &    (insertion)\\
ms_{s,t}(i-k,j-1)+\delta(\alpha_{i-k+1}\dots \alpha_i\rightarrow \beta_j) &    (consolidation)\\
ms_{s,t}(i-1,j-k)+\delta(\alpha_i\rightarrow \beta_{b-k+1}\dots \beta_{j}) &    (fragmentation)\\
                 \end{cases*} 
\]
                 
Here it can be seen two new operations \textbf{Consolidation} and \textbf{Fragmentation} are introduced. A summary of these operations is provided below. For more information refer to  \cite{mongeau1} or  \cite{mongeau2}

\paragraph{Consolidation}
This operation compresses a few characters into a single one. For example, two 16th notes can be clubbed together to give an 8th note. 

\paragraph{Fragmentation}
This operation is the opposite of consolidation. It involves splitting a larger note quantity into smaller units.
compresses a few characters into a single one. For example, two 16th notes can be clubbed together to give an 8th note. 
\newpage
\begin{figure}[H]
	\centering
	\includegraphics[width=0.6\textwidth]{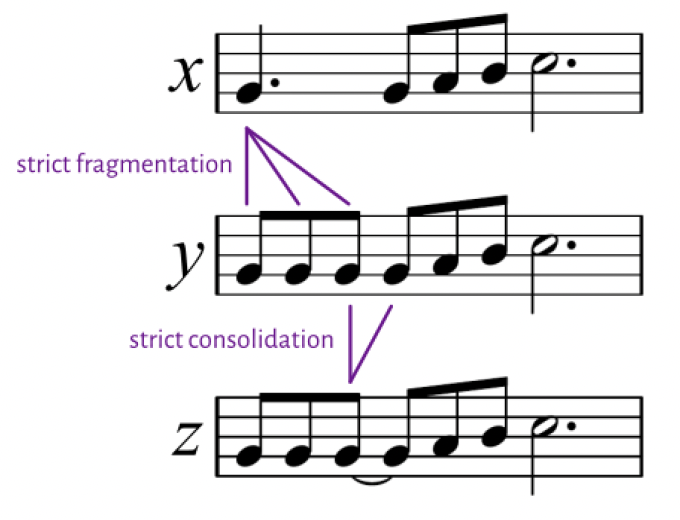}
				\caption{Fragmentation and Consolidation  \cite{mongeau2}-In the picture given above the\textbf{ dotted note} is equivalent to \textbf{three notes}. Fragmentation involves splitting that note into three notes. Consolidation involves combining two notes(the line between two tied notes).}
\end{figure}
\textbf{Assumptions and Issues}\\
By construction the Mongeau-Sankoff algorithm relies on certain features being known beforehand.
\begin{enumerate}
\item \textbf{Tempo}-The Mongeau-Sankoff algorithm requires both \textbf{Note values} \emph{and} the corresponding \textbf{Note Durations}. Here note durations are in \textbf{music units} (16th notes,8th notes etc.). This requires the tempo of the song to be known. We assume 16th notes as the basic unit. Note duration can then be converted from seconds according to the formula below
\begin{equation}
    \boxed{Dur(N)=\frac{4*Tempo}{15}\cdot (Dur_s(N))}
\end{equation}

where $Dur(N)$ is the note duration in music units, $Dur_s(N)$ is the note duration in seconds, Tempo is the standard tempo value in beats per minute. 

While accounting for Note Durations is a good feature on paper, the issue is that the algorithm deals in music units. This means the tempo of every song needs to be known. This is not possible for most data since the tempo values store in midi files are often default values (Tempo=120). We found a workaround this issue by computing the tempo of each song through a Tempo detection algorithm( \cite{tempo}). Of course, this value is only an approximation and will thus lead to poor results.

\item \textbf{Scale Invariance}- In most cases songs are written in a certain Key (eg. Cmaj, Cmin etc.) with its corresponding scale. We can transpose from one scale to another (major$\rightarrow$major, minor$\rightarrow$minor) while communicating the same musical idea (melody). Thus, it is possible the input midi might be in a different scale compared to our database and will lead to poor matches. This is why we transpose all major scales to Cmaj and all minor scales to Cmin. Thus, the same song can be recognised even if it is presented in a different scale. 

Again, this seems like a good idea-and is something we thought of independently as well. However, this runs into major problems. When we transpose data into one scale (or equivalently, into 12 notes) we end up with a \textbf{loss of information}. Given a large enough database, the number of collisions(having same or very similar note representations) will increase substantially. We analyse these results in our experimentation section

\item \textbf{Length Similarity} Mongeau Sankoff assumes that the length of both strings are of the same length. This is combated by using sliding windows as discussed in the two RSA algorithms above.
\end{enumerate}
\newpage
\subsection{Extensions to The Model}

 \subsubsection{V.S.M. Approach}
    Another approach to retrieval could be using a modified Vector Space Model.Here the vector 
    We can create a vector space model where the indices will be decided by the Note instead of the words in vocabulary and the weights associated with each index will be a basis function of the variables time duration and the velocity of the Note,
    \[ V[j] = \phi (vel_j,duration_j) \]
    \vspace{-5pt}
    \[ Sim_{cosine} (m,q) = \frac{\vec{m} . \vec{q}}{\|\vec{m}\| \|\vec{q}\|} \]
    \\here, $j$ represents the Note index, vel represents the velocity of the Note and duration represents the \textbf{total duration} the note is played in the song. This is equivalent to the \textbf{term frequency} parameter in VSM for text documents. 
    
    The pitch can lie between (-8192, 8191) so we will have a 2*8192 dimensional vector for each song snippet representation with the weights corresponding to some basis function of the velocity and the time duration while that pitch was on during the complete audio.  

    This process has a major con as it disregards the order of occurrence of the notes in the given 5 second clips, hence it can't be directly used for retrieval however, it \textbf{can be used for trimming down our documents to search} as VSM based similarity can be computed quickly and if it is above some threshold, only then we look into the edit distance similarity computation.
    
    \subsubsection{Extension to Dejavu}
    
    The approach followed in methods like Dejavu  \cite{dejavu} is that they create a spectrogram of the original song by using FFT over small windows. Once they have spectrogram, they proceed to the next step of finding peeks (local maxima of amplitudes) in that spectrogram. Now, the data left to store is the position of these maximas (only frequency and time required). These discrete (time,frequency) pairs are in theory resistant to noise. To create these fingerprints \cite{chang_neural_2021}, they store these peaks parametrized by its neighbours (time difference between nearest k peaks) by passing this information to a hash function and storing the output (SHA1). The value of k (termed as Fan Value), is a hyper-parameter and controls how many peaks should a particular peak be associated with before passing to the hash function. A higher fan value will lead to more number of fingerprints but better accuracy.
    
    Dejavu suffers from a problem where if we change the speed of playing the notes in the query, then the system wont be able to recognize this as its time dependent. Our Text-based similarity approach should theoretically outperform Dejavu on the basis of metrics such as Recall etc.

\section{Experiments}
We evaluate the performance of Clarinet over the \textbf{Maestro} Dataset  \cite{hawthorne2018enabling} with the metrics as defined.
\begin{definition}[\textbf{Recall@K}]
\emph{The definition of Recall@K is as one would assume intuitively. It is the recall of the model assuming the correct audio (or document) file is present in the top K when similarity scores using a certain method are sorted in descending order.}
\end{definition}

 \begin{definition}[\textbf{Mean Reciprocal Rank, MRR}]
 \emph{Mean reciprocal rank of the desired document over the queries. \textbf{MRR only cares about the single highest-ranked relevant item}. Higher ranks are penalised by decreasing MRR.}
\[MRR=\frac{1}{Q} \sum_{q=1}^{Q}\frac{1}{rank_q}\]
\emph{where $rank_q$ is the rank of the document from where the query was sampled.}
\end{definition}
\begin{definition}[\textbf{Margin of Discrimination(MD)}]
\emph{Difference between similarity scores of the desired document with the next document. Measures how well the system could discriminate between the document from where the query was sampled and other documents.}
\end{definition}
 \begin{definition}[\textbf{Time Taken}]
 \emph{The average time taken per query is recorded and compared.}
 \end{definition}
\newpage

\subsection{Dataset}
The original dataset we chose (MAESTRO dataset) had 1287 songs with on average every song being roughly 20 minutes in length. These were originally recordings from ten years of International Piano-e-Competition. 
\\
\textbf{Documents}
We generate the documents by clipping the original dataset into \textbf{20 seconds} each from the beginning. The reason behind doing so was so as to increase the retrieval time as the songs in this dataset were huge(~10 minutes).
\\
\textbf{Queries}
As there weren't any queries available on the internet for this task, we created the queries ourselves by running a script which randomly selects 40 songs from our dataset and then randomly clips \textbf{5 seconds} from anywhere in between. The start time need not be in multiple of seconds and can start from anywhere. 


\subsection{Melody Creation}
Creating the melody from the MIDI audio file was done via the skyline algorithm and its modified version. 
Both the methods resulted in a MIDI file that we could use our retrieval models on, with the modified version doing substantially better as seen from the results of the experimentation as seen below which contains the MIDI representation of the files.
\begin{figure}[H]
	\centering
	\includegraphics[width=.5\textwidth]{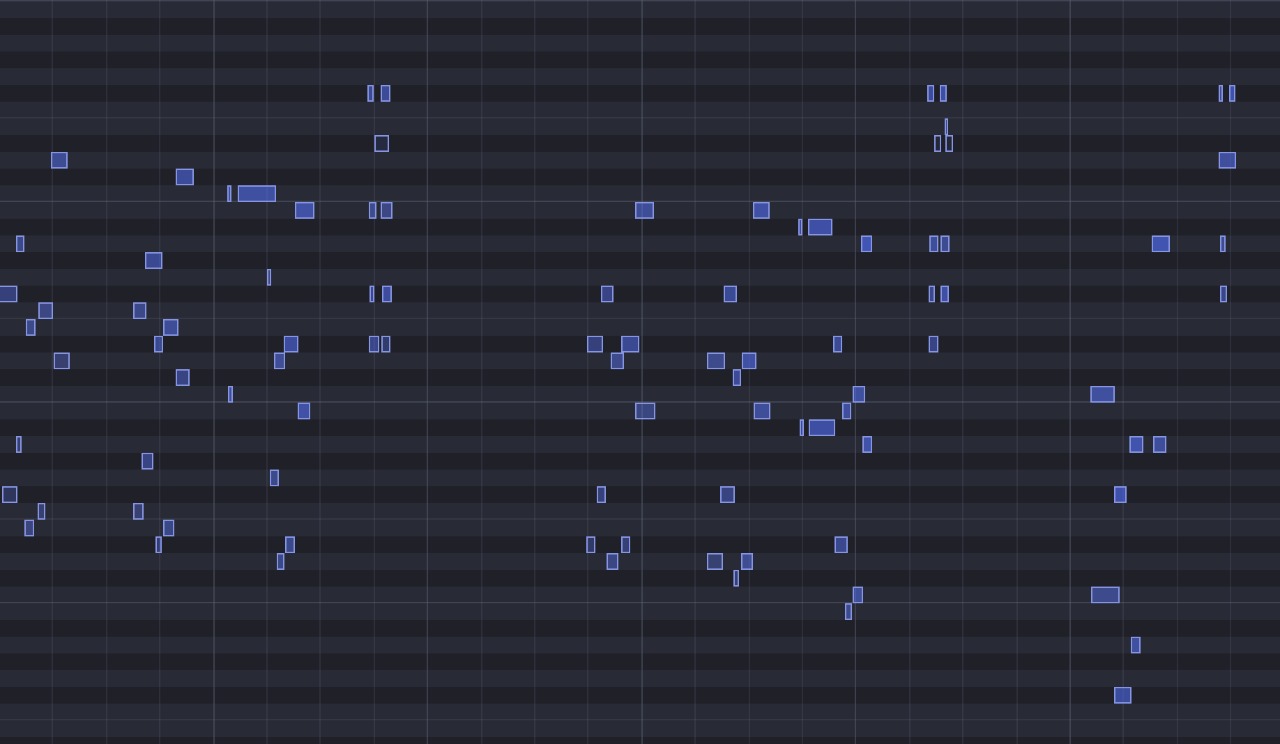}
				\caption{MIDI Representation of original audio}
\end{figure}
\begin{figure}[H]
	\centering
	\includegraphics[width=.5\textwidth]{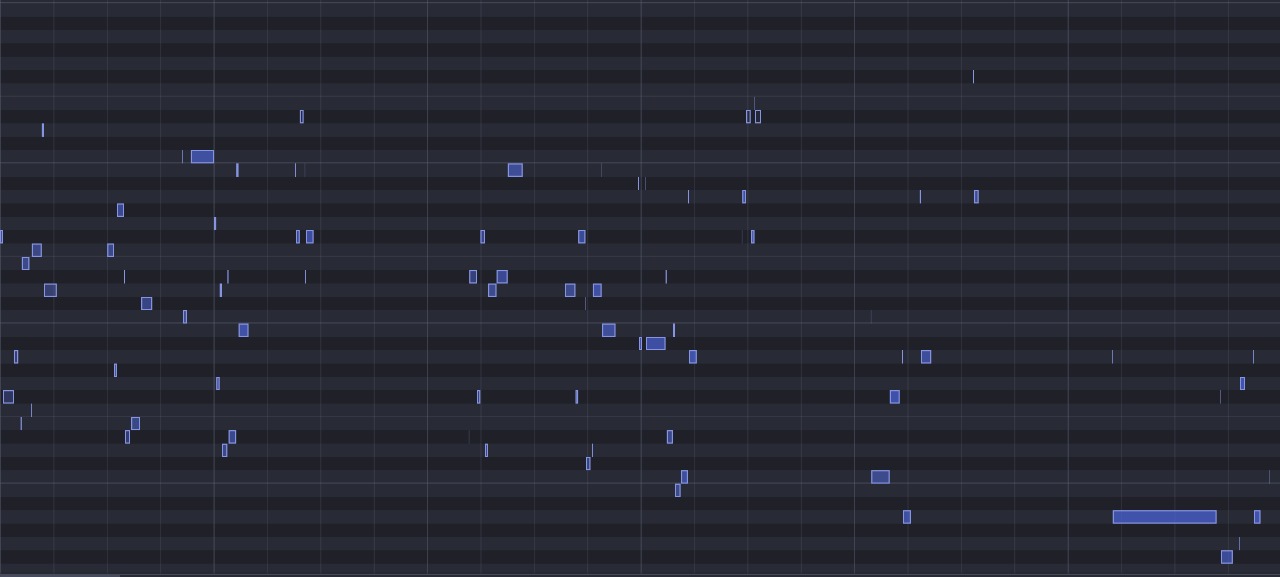}
				\caption{MIDI Representation of skyline extracted melody}
\end{figure}
\begin{figure}[H]
	\centering
	\includegraphics[width=.5\textwidth]{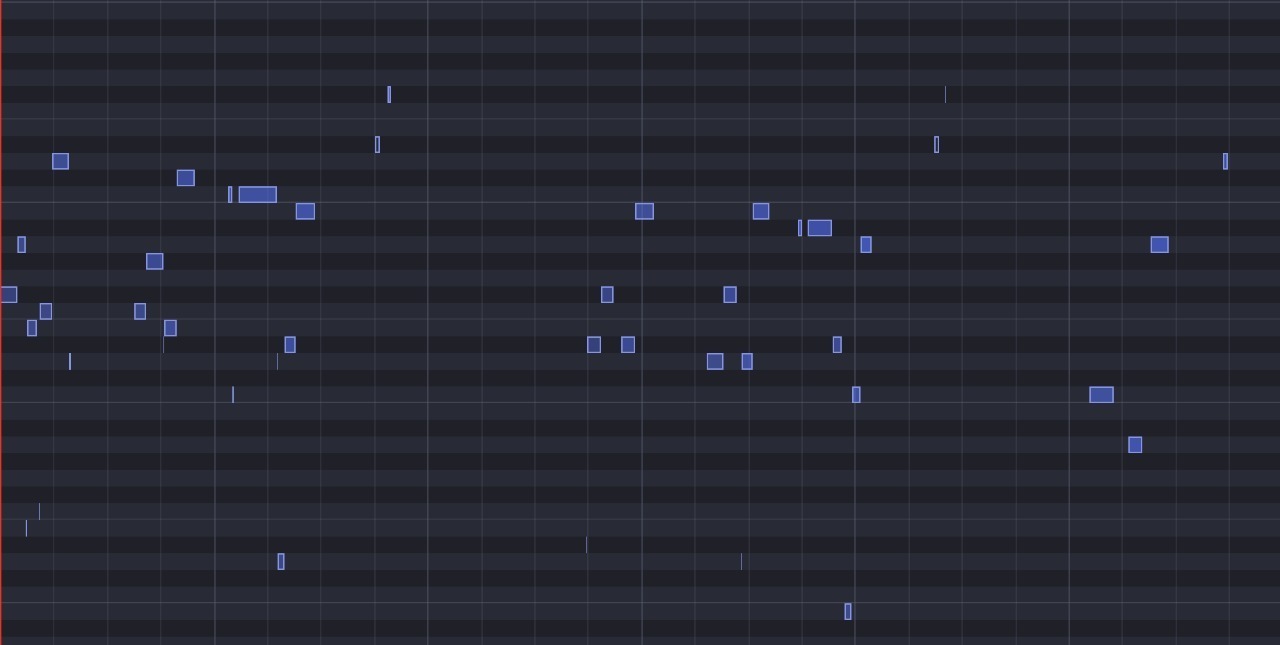}
				\caption{MIDI Representation of modified skyline extracted melody}
\end{figure}
\newpage
Hence we can see that the modified skyline algorithm is cleaner and stable\footnote{as defined by \textbf{Definition 3.1}}. While we have not shown any numeric metric for comparison here (we do so in the next section), it is clear from the data representation that our modified algorithm surpasses the existing state of the art method for extracting melodies by a large margin.

\subsection{Retrieval Evaluation}
We will now utilise this section to showcase our results and the improvements on existing state of the art methods utilising our own modifications or novel methods.
\\The first metric we will be showcasing is Recall@K for $K=\{1,3,5,10\}$.
\subsubsection{Recall@K}

\begin{table}[H]
\centering
\begin{adjustbox}{width={\textwidth},totalheight={\textheight},keepaspectratio}
\begin{tabular}{|
>{\columncolor[HTML]{FFFFC7}}c |
>{\columncolor[HTML]{FFFFC7}}c |
>{\columncolor[HTML]{FFFFC7}}c |
>{\columncolor[HTML]{FFFFC7}}c |
>{\columncolor[HTML]{FFFFC7}}c |
>{\columncolor[HTML]{FFFFC7}}c |
>{\columncolor[HTML]{FFFFC7}}c |}
\hline
\textbf{Similarity Method} & \textbf{Melody Extractor} & \textbf{Processed} & \textbf{Recall@1} & \textbf{Recall@3} & \textbf{Recall@5} & \textbf{Recall@10} \\ \hline
RSA Note                       & Modified                  & Unprocessed        & 1.0               & 1.0               & 1.0               & 1.0                \\ \hline
RSA Note                       & Skyline                   & Unprocessed        & 1.0               & 1.0               & 1.0               & 1.0                \\ \hline
RSA Time                       & Modified                  & Unprocessed        & 0.947             & 0.973             & 0.9736            & 1.0                \\ \hline
RSA Time                       & Skyline                   & Unprocessed        & 0.842             & 0.947             & 0.973             & 0.97368            \\ \hline
RSA Note                       & Skyline                   & Processed          & 0.526             & 0.657             & 0.684             & 0.6842             \\ \hline
RSA Note                       & Modified                  & Processed          & 0.473             & 0.60              & 0.631             & 0.657              \\ \hline
RSA Time                       & Modified                  & Processed          & 0.0526            & 0.07              & 0.1578            & 0.236              \\ \hline
RSA Time                       & Skyline                   & Processed          & 0.0526            & 0.157             & 0.210             & 0.236              \\ \hline
\end{tabular}
\end{adjustbox}
\vspace{0.8em}

\caption{Recall Comparison}
\label{tab:recall-table}
\end{table}

\begin{figure*}[h]
    \centering
    \begin{subfigure}[t]{0.49\textwidth}
        \centering
        \includegraphics[scale=0.4]{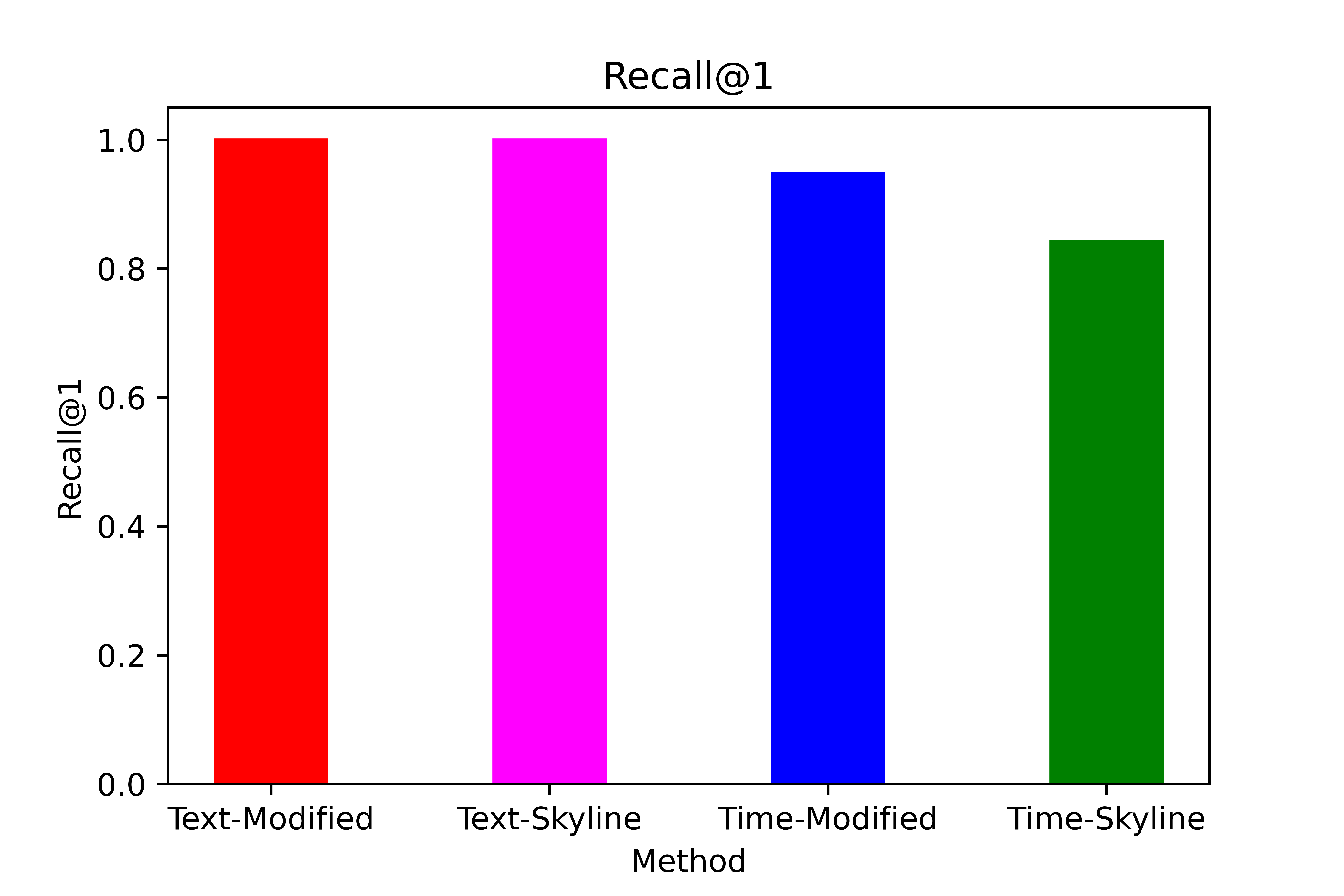}
 				\caption{Recall@1}
    \end{subfigure}%
    ~ 
    \begin{subfigure}[t]{0.49\textwidth}
        \centering
        \includegraphics[scale=0.4]{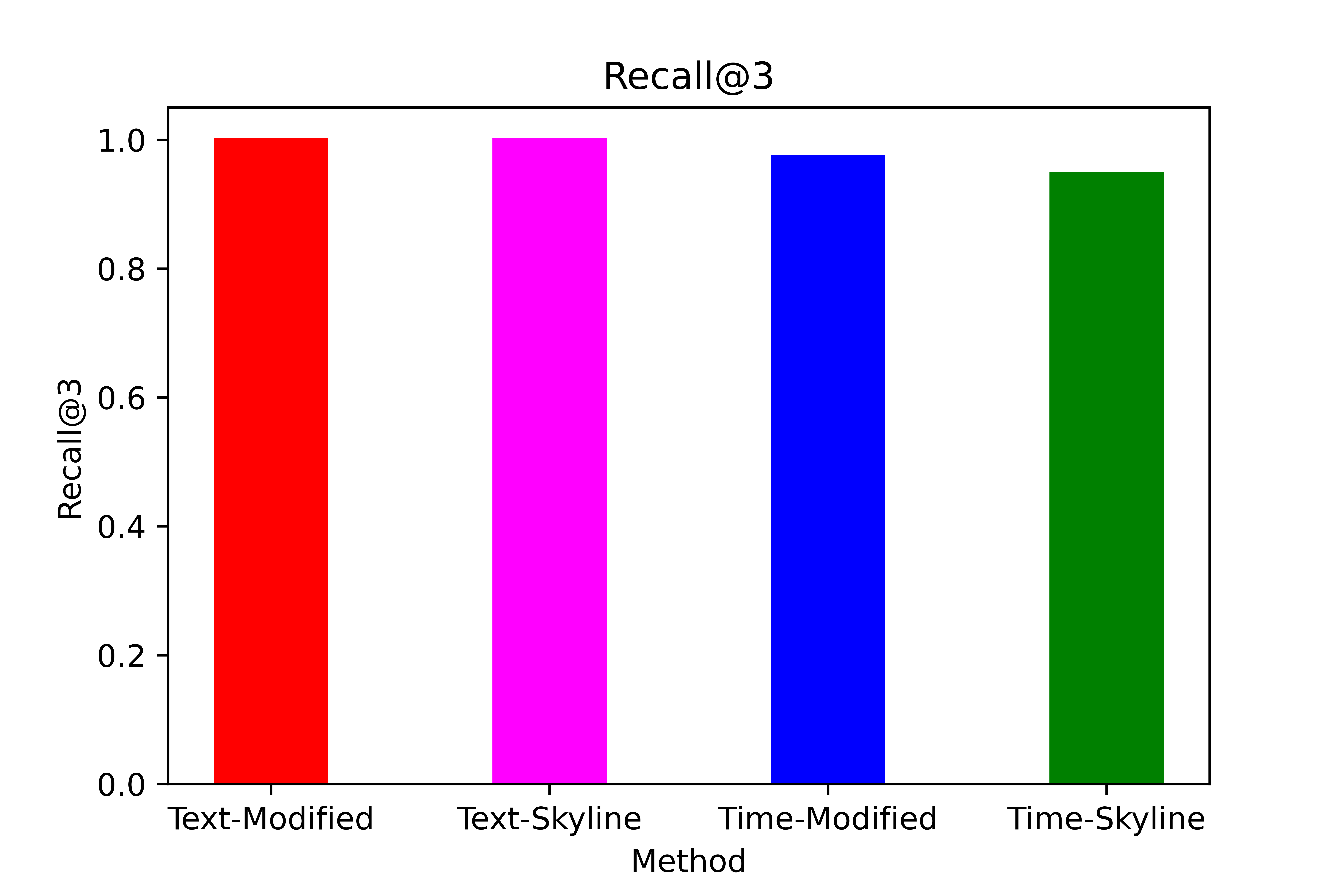}
 				\caption{Recall@3}
    \end{subfigure}
    \caption{Recall@K for K = 1, 3}
\end{figure*}
\begin{figure*}[h]
    \centering
    \begin{subfigure}[t]{0.49\textwidth}
        \centering
        \includegraphics[scale=0.4]{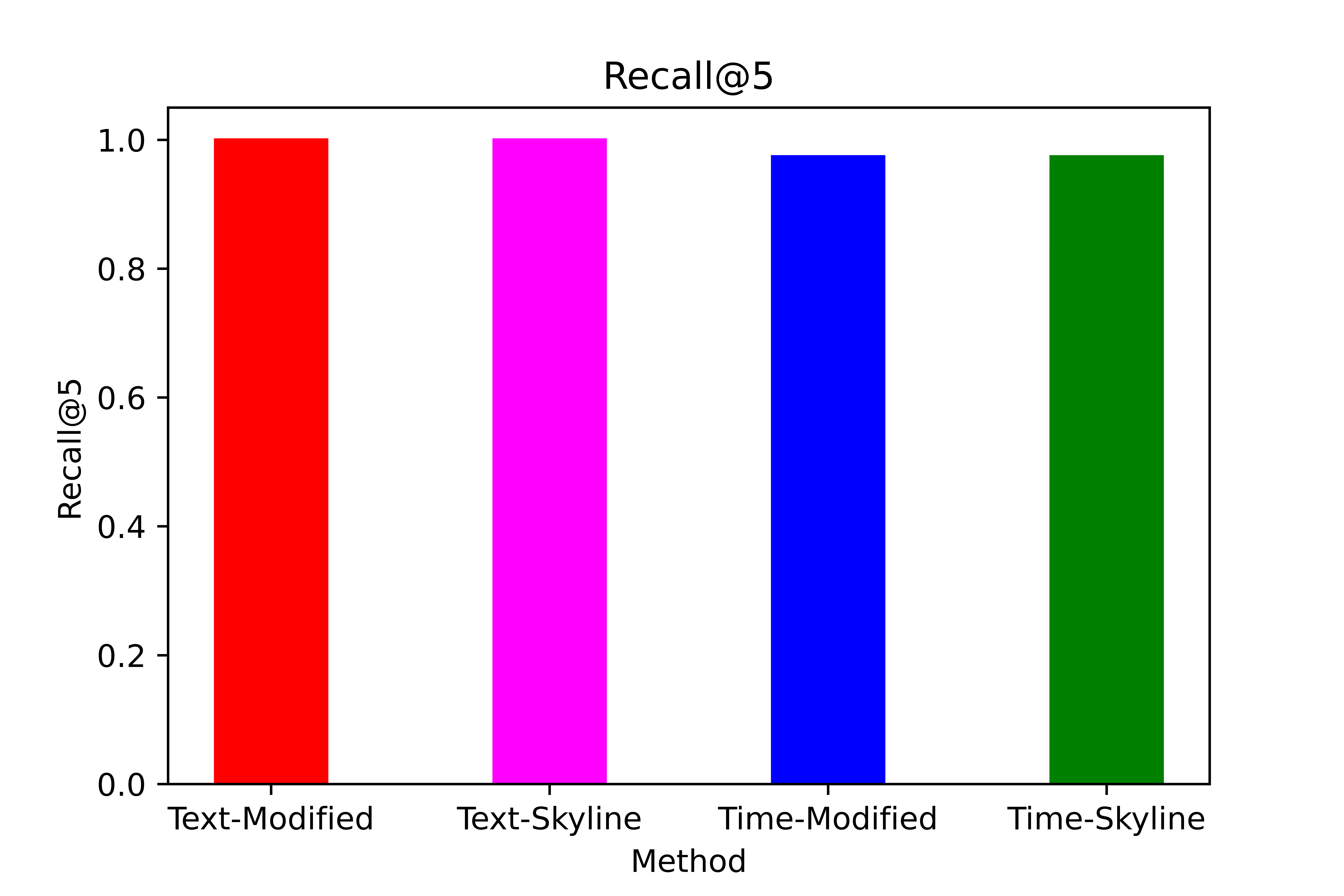}
 				\caption{Recall@5}
    \end{subfigure}%
    ~ 
    \begin{subfigure}[t]{0.49\textwidth}
        \centering
        \includegraphics[scale=0.4]{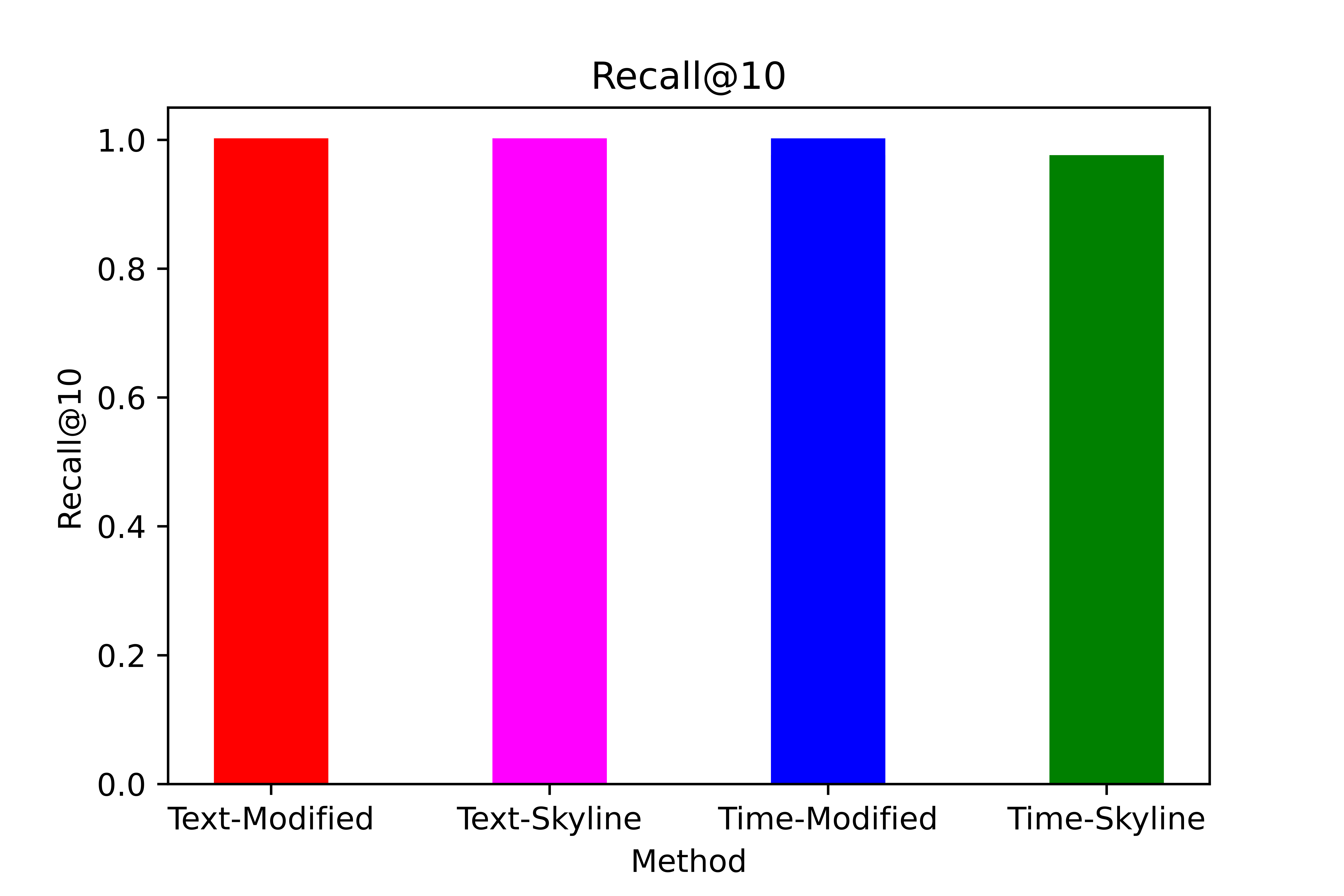}
 				\caption{Recall@10}
    \end{subfigure}
    \caption{Recall@K for K = 5, 10}
\end{figure*}
\newpage
 \begin{figure}[H]
 	\centering
 	\includegraphics[scale=0.8]{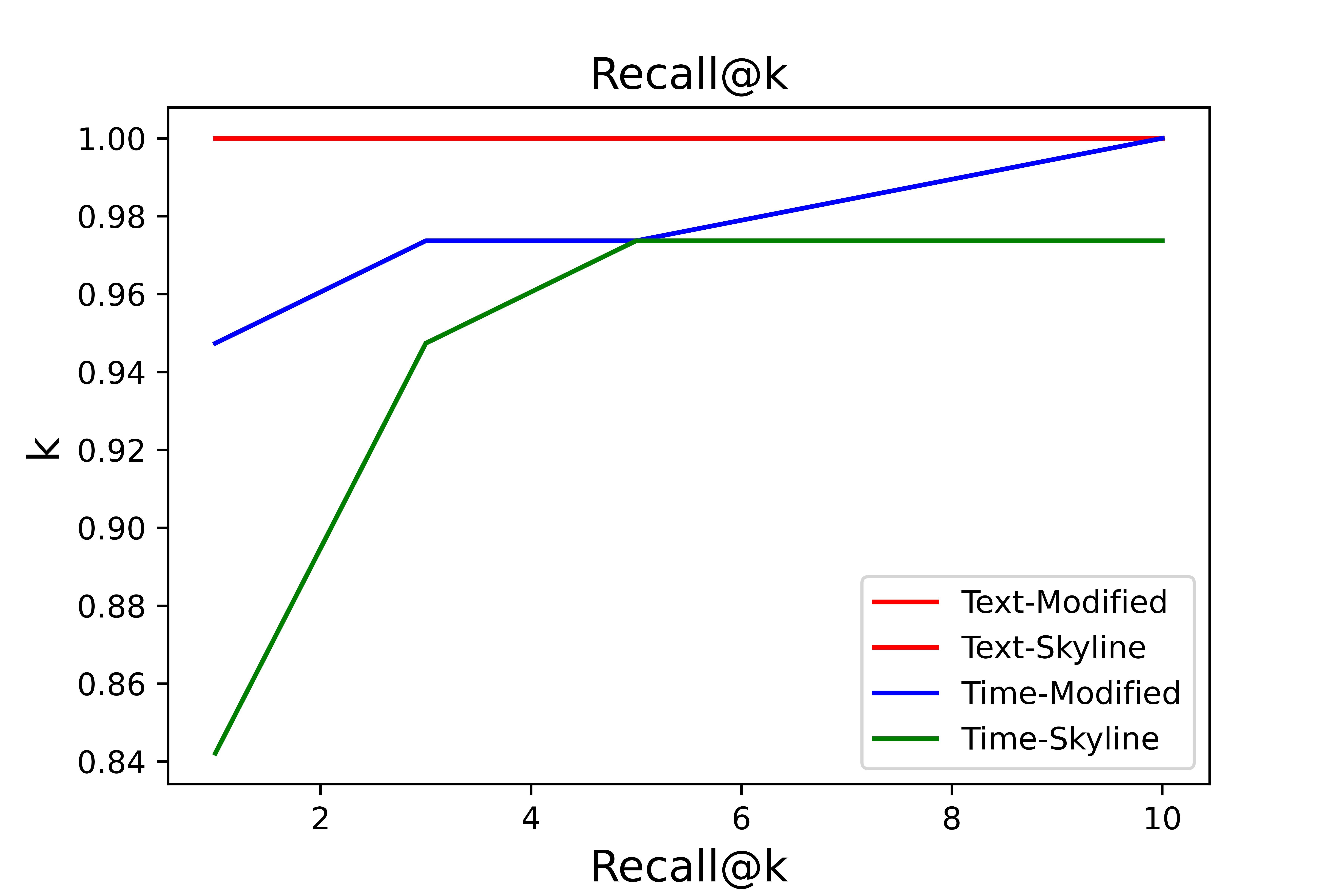}
 				\caption{Recall@K vs K}
 \end{figure}
 
From these results, we can make the following observations:

\begin{enumerate}
    \item \textbf{RSA Text based methods reach 100\% recall@1}, i.e. always succeed in predicting the correct result at the first position. This happens because it matches the query and document by looping through windows of length = query length. 
    \item The \textbf{Modified Skyline method outperforms} the classical Skyline method in both methods.
    \item The \textbf{RSA Time} method also has a decent recall@1 and mostly achieves \textbf{100\% recall@3}, i.e. the correct result lies in the first 3 documents always. (RSA Time - Skyline).
    \item \textbf{Effect of Processing}-Processing increases time of retrieval substantially but leads to \textbf{unacceptablly bad results}. This is due to low dimensional data. It can be seen using \textbf{RSA Time} on Processed queries led to results being randomised. This is because RSA Time is susceptible to noise and Processing queries makes our model \textbf{robustness poor}. 
\end{enumerate}
\emph{For the analysis below we exclude Processed queries due to very poor retrieval quality.}
\subsubsection{Mean Reciprocal Rank (MRR)}

\begin{table}[H]
\centering
\begin{tabular}{|
>{\columncolor[HTML]{FFFFC7}}c |
>{\columncolor[HTML]{FFFFC7}}c |
>{\columncolor[HTML]{FFFFC7}}c |}
\hline
\textbf{Similarity Method} & \textbf{Melody Extractor} & \textbf{Mean Reciprocal Rank (MRR)} \\ \hline
RSA Note                       & Modified                  & 1.0                                 \\ \hline
RSA Note                       & Skyline                   & 1.0                                 \\ \hline
RSA Time                       & Modified                  & 0.964                               \\ \hline
RSA Time                       & Skyline                   & 0.900                               \\ \hline
\end{tabular}
\vspace{0.8em}
\caption{Mean Reciprocal Rank Comparison}
\label{tab:mrr-table}
\end{table}
The mean reciprocal rank was found for all combinations. This metric has similar results as recall. RSA note performed better than RSA Time and Modified Skyline performed better than traditional Skyline.
\newpage

\subsubsection{Margin of Discrimination}
\begin{table}[H]
\centering
\begin{adjustbox}{width={\textwidth},totalheight={\textheight},keepaspectratio}
\begin{tabular}{|
>{\columncolor[HTML]{FFFFC7}}c |
>{\columncolor[HTML]{FFFFC7}}c |
>{\columncolor[HTML]{FFFFC7}}c |
>{\columncolor[HTML]{FFFFC7}}c |
>{\columncolor[HTML]{FFFFC7}}c |}
\hline
\textbf{Similarity Method} & \textbf{Melody Extractor} & \textbf{Processed} & \textbf{Recall@1}  & \textbf{\begin{tabular}[c]{@{}c@{}}Margin of Discrimination\\ (MD)\end{tabular}} \\ \hline
RSA Note                       & Modified                  & Unprocessed        & 1.0                & 28.5\%                                                                           \\ \hline
RSA Note                       & Skyline                   & Unprocessed        & 1.0                & 29.09\%                                                                          \\ \hline
RSA Time                       & Modified                  & Unprocessed        & 0.947 & 19.10\%                                                                          \\ \hline
RSA Time                       & Skyline                   & Unprocessed        & 0.842  & 20.27\%                                                                          \\ \hline
\end{tabular}
\end{adjustbox}
\vspace{0.8em}
\caption{Margin of Discrimination Comparison}
\label{tab:md-table}
\end{table}

Margin of Discrimination is the difference in confidence scores(similarity) between the similarity made for the target document and the similarity of the document just after the target document in ranking. These scores also normalized. This is an indicative of how \textbf{confident the model is in discriminating} its prediction from the other documents. 

We observe that\textbf{ RSA Text based approaches get the highest Margin of Discrimination} followed by the RSA Time approach. The choice of melody extraction algorithm seems to be playing little role here.

\subsubsection{Time Taken}
\begin{table}[H]
\centering
\begin{tabular}{|
>{\columncolor[HTML]{FFFFC7}}l |
>{\columncolor[HTML]{FFFFC7}}l |
>{\columncolor[HTML]{FFFFC7}}l |}
\hline
\textbf{Similarity Method} & \textbf{Melody Extractor} & \textbf{Time per query (sec)}      \\ \hline
RSA Time                       & Modified                  & 29.17 \\ \hline
RSA Time                       & Skyline                   & 62.91  \\ \hline
RSA Note                      & Modified                  & 183.86 \\ \hline
RSA Note                       & Skyline                   & 572.43  \\ \hline
\end{tabular}
\vspace{0.8em}
\caption{Time per query}
\label{tab:time-table}
\end{table}

We observe that there's a \textbf{trade-off between recall scores and the time taken} to answer every query. Although RSA Text approach gets 100\% recall@1, it has a major drawback of \textbf{increased retrieval time}. RSA Time approach still gets very decent results with significantly lower retrieval time. 

Moreover, we also see that \textbf{Modified Skyline approach outperforms the classical Skyline} approach in terms of time taken per query as well.
\newpage
\section{Conclusion}
This paper contains an entire pipeline for building a robust music retrieval system, outlining various state of the art methods for melody extraction and similarity functions, our modifications of them and three novel algorithms in the same domain. 
\\The novel algorithms are listed below:
\begin{itemize}
    \item \textbf{Modified skyline algorithm}: It clearly outperformed the original skyline algorithm in a music retrieval task that is generally robust to quality of melody extractor! We could see an improvement in various other parameters as well like stability, flexibility and retention of additional information.
    \item \textbf{RCA Time and Note}: These novel algorithms in music retrieval performed extremely well benchmarked against each other and the modified Mongeau-Sankoff algorithm, achieving a recall of 100\% when paired with the novel melody extractor.
\end{itemize}
Alongside these novel algorithms, we also modified the Mongeau-Sankoff algorithm to utilise it for music retrieval by changing the following:
\begin{itemize}
    \item \textbf{Tempo Invariance}: Mongeau-Sankoff requires the tempo to be known beforehand which is not always known to us. Thus we implemented a technique that will auto detect the tempo of the audio, and thus allows us to use Mongeau-Sankoff without knowing tempo beforehand.
    \item \textbf{Scale Invariance}: We transposed all scales of the audio to Cmajor (or Cminor depending on original scale) to make the data scale invariant, which allows the Mongeau-Sankoff algorithm to work in general as well.
    \item \textbf{Length Similarity}: Mongeau-Sankoff assumes the length of the files to be same. This is rarely the case, and hence we implemented a sliding window technique to ensure Mongeau-Sankoff can work for general audio files as well.
\end{itemize}
The end conclusion of all this work entails that we built a music retrieval system that had the following improvements and properties:
\begin{itemize}
\item \textbf{Robust}: Due to the melody extraction being significantly improved, our data is more robust to noise that may creep in. This is because it will be filtered out by the melody extractor, and won't disturb our retrieval system at all.
\item \textbf{Discriminatory}: As seen by the experimentation, our model gains a high margin of discrimination and thus improves the confidence that our model will work well even under exceptional circumstances. 
\item \textbf{General}: As mentioned above, our model is invariant to a lot of variations in the input. That allows us to utilise our model on a wide range of data, making it extremely general.
\item \textbf{Speed}: Our model pre-processes data in a way that allows for a huge speed bump. Coupled with the fast retrieval mode (RSA Time), we can achieve high speeds to retrieve audio files.
\item \textbf{Accurate}: Finally, the most important aspect of any model, our retrieval system is extremely accurate. Using the proposed novel model(s), we get a 100\% accuracy while also being extremely fast!
\end{itemize}
\newpage
\section{Future Work}
\begin{itemize}
    \item \textbf{Stride Length Analysis}: As talked about previously, both RSA algorithms rely on sliding windows. We know accuracy decreases with increase in stride length-but by how much? Increasing stride length would cause fewer number of notes to be considered and would thus be faster. This is thus a \textbf{hyperparameter} which we can tune according to our data. 
    
    \item \textbf{Humming implementation}: We plan on testing the dataset on real world scenario by singing tones such as Happy birthday into the microphone and then converting the recorded WAV file into a MIDI file which can be further used as a query.
    \item \textbf{Composer detection}: What is usually done for these tasks is that they perform detection by feeding the spectrograms of the songs to a classifier  \cite{composer}. Our IR implementation can also be used for this task by picking out the most common composers with high similarity scores and giving that as the output.  
    \item \textbf{Compare to DejaVu}: This is one scope of improving the evaluation we had so far. Dejavu suffers from a problem where if we change the speed of playing the notes in the query, then the system wont be able to recognize this as its time dependent. Our Text-based similarity approach should theoretically outperform Dejavu on the basis of metrics such as Recall etc.
    \item \textbf{Vector Space Models assistance}: We can trim down on the searches by first performing a VSM based similarity of our queries with the data. If the similarity is above some threshold, only then we move on to comparing the similarities on the basis of our Levenshtein distance (costly operation). 
    \item \textbf{Edit distance improvement through R* trees} : The current implementation of edit distance is of the complexity O(mxn) where m and n are the lengths of the note representations. This can be further improved to O(mlogn) by using R* trees. Moreover, the current implementation is not vectorized as well and this is something which can be done to improve the retrieval speeds.  \cite{Rtrees}.
    
\end{itemize}

\section{Possible Applications}
\begin{enumerate}
\item \textbf{Music Recognition}-The most obvious application of our project is in music recognition. The user can input a song derivative as their query and get back the song from which the sample was excerpted-thus working as an extended version of Shazam
\item \textbf{Melodic Similarity}-Suppose a musician writes a melody. He could send this melody as a query and get back a list of results of pre-existing songs that sound similar. He could then use ideas from these songs to continue his song.
\item \textbf{Copyright Infringement Detection}-Since the project  is based essentially on melody matching, it could also be used on platforms like Youtube where detecting whether a video has used a copyrighted song becomes important
\end{enumerate}

\newpage 
\pagenumbering{roman} 

\bibliography{references.bib} 
\bibliographystyle{plain} 

\end{document}